\documentclass[aps,pra,preprint,superscriptaddress]{revtex4}

\usepackage{graphicx}
\usepackage{epstopdf}
\usepackage{dcolumn}
\usepackage{bm}

\begin{document}

\title{Decoherence Free Neutron Interferometry}

\author{D. A. Pushin}
\email[]{mitja@mit.edu}

\affiliation{Massachusetts Institute of Technology, Cambridge, Massachusetts 02139, USA}
\author{M. Arif}
\affiliation{National Institute of Standards and Technology, Gaithersburg, Maryland 20899, USA}
\author{D. G. Cory}
\affiliation{Massachusetts Institute of Technology, Cambridge, Massachusetts 02139, USA}

\date{September 10, 2008}

\begin{abstract}
Perfect single-crystal neutron interferometers are adversely sensitive to environmental disturbances, particularly mechanical vibrations. The sensitivity to vibrations results from the slow velocity of thermal neutrons and the long measurement time that are encountered in a typical experiment. Consequently, to achieve a good interference solutions for reducing vibration other than those normally used in optical experiments must be explored. Here we introduce a geometry for a neutron interferometer that is less sensitive to low-frequency vibrations. This design may be compared with both dynamical decoupling methods and decoherence-free subspaces that are described in quantum information processing. By removing the need for bulky vibration isolation setups, this design will make it easier to adopt neutron interferometry to a wide range of applications and increase its sensitivity.
\end{abstract}

\maketitle

\section{Introduction}

Neutron interferometry is one of the most precise techniques used to test the postulates
of quantum mechanics,  and it is also one of the most important and precise techniques used to measure low-energy neutron cross sections \cite{ni_book}. Although the fundamentals of neutron
interferometry are easily recognizable from common optics, the
slow velocity of neutrons (1680 $m/s$ for the 2.35 $\AA$ neutrons
used in this study) and low count rates at the
detector ($1000/min$) demand novel solutions.  The most important
of these was the development of multiblade, single-crystal
interferometery which enables high contrast to be observed with only
limited beam alignment \cite{rauch1int1974,bauspiess1int1974}.  The
challenge remains however of making the experiment robust against
mechanical vibrations \cite{shull.vibrations.book,arif_inter}.  A
typical single-crystal interferometer has path lengths of 10 $cm$
with a typical 50 $\mu s$ travel time of neutrons in the system.
  Small-amplitude low-frequency vibrations may significantly degrade the contrast.   Here we propose a
solution to robust interferometry based on a  four-blade single-crystal
geometry that reduces errors introduced by vibrations.

\section{Interferometer Schematic}

\begin{figure}[!ht]
\includegraphics[width=1\textwidth]{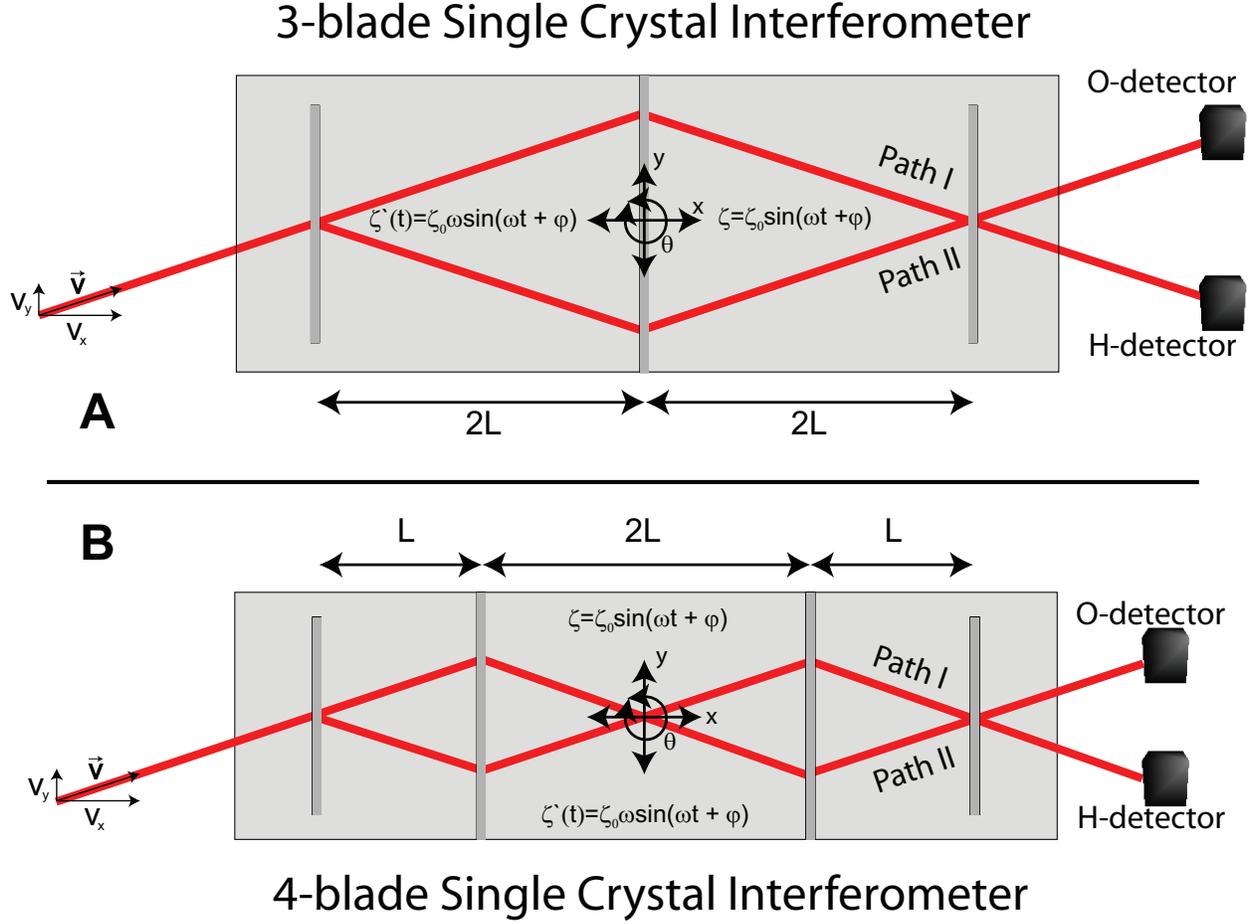}
\caption{\label{fig:4bl_sch} \textbf{A}: A schematic diagram of the
three-blade neutron interferometer. A neutron beam (with neutron
velocity $\vec{v}$) comes from the left, is split by the first
blade, is diffracted on the second blade, and recombines at the
third. After passing through the interferometer, the beam is
captured by the O- and H-detectors. We model vibrations as
oscillations of and around the center of mass of the interferometer,
as $\zeta(t)=\zeta_0\sin{(\omega t + \varphi)}$, where $\zeta$ could
be $y$ (transverse vibrations), $x$ (longitudinal vibrtions), and $\theta$ 
(rotational vibrations). In order to compare oscillations between the three- and four-blades
devices the distance between the blades is set equal to $2L$.
\textbf{B}: A schematic diagram of the proposed interferometer with
four blades. Instead of one diffracting blade here we have two, which
reverses the neutron paths in order to compensate for vibrations. We will compare the performance
using the same vibration modes with the same amplitudes. Note that for the three-blade interferometer the O-detector has the maximum contrast and in the four-blade interferometer the H-detector has the maximum contrast.}
\end{figure}

The most common three-blade geometry for a perfect crystal neutron interferometer (NI)~\cite{ni_book}
is show in Fig.~\ref{fig:4bl_sch}~\textbf{A}) along with the four-blade
configuration (Fig.~\ref{fig:4bl_sch}~\textbf{B}) that is less
sensitive to low-frequency vibrations.

In the three-blade case the neutron beam, coming
from the left, is coherently split into two paths by the first blade
via Bragg scattering. After being reflected by the second blade,
these two paths are recombined at the third blade. The resultant
interference is observed at the O- and H-detectors. Note that we
align the $y$ coordinate parallel to the Bragg planes and the $x$
coordinate is perpendicular to the Bragg planes. In the four-blade case
the situation is nearly identical with the significant difference
that the paths are reflected and cross each other without mixing
at the center of the interferometer, i.e. without a blade there is no mixing of the states.

\begin{figure}[!ht]
\centering
\includegraphics[width=0.7\textwidth]{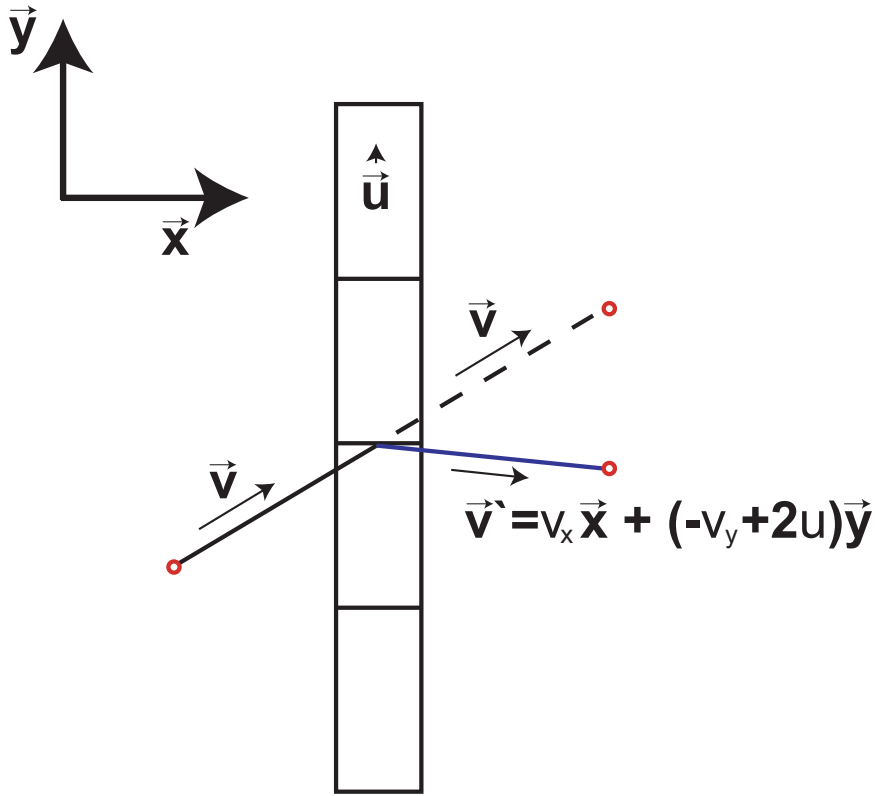}
\caption{\label{fig:inter_sch} A schematic diagram of the neutron
scattering from a blade of the neutron interferometer. Due to the crystal movement, the
reflected neutron's momentum is modified. For the transmitted
case, the momentum remains unaltered.}
\end{figure}

It is sufficient to take a simple model for noise and to consider vibrations as sinusoidal oscillations around the center of mass of the single crystal, which we write as
\begin{equation}
\zeta(t)=\zeta_0\sin{(\omega t + \varphi)} \label{eq:oscil}.
\end{equation}
For $\zeta$ we can specify any coordinates $x$, $y$, $z$, or any angles
(such as $\theta$ - rotation around z axis). In order to motivate
the discussion we adopt a simple model for the neutron-blade
interaction that includes all of the necessary physics. The interaction
is that of bouncing of a small
particle (neutron) from a moving heavy wall (the blade) where the
particle is reflected. When the particle is transmitted there is no
interaction (figure~\ref{fig:inter_sch}). We use conservation of
momentum and energy to calculate the neutron's change in velocity
after bouncing. We require a small enough amplitude and low enough
oscillation frequency that the modified momentum of the neutron
still satisfies the Bragg condition.

Vibrations thus modify the neutron velocities and change the
path length of the neutron inside the interferometer \cite{shull.moving.xtal:678,buras.moving.lattices,dresden.sagnac.effect.PhysRevD.20.1846,mashhoon.PhysRevLett.61.2639}.

Using these approximations, it is clear that vibrations along every
axis except the $z$ coordinate will reduce the interferometer
contrast. The $z$-component of neutron velocity is zero and the paths
lengths are independent of crystal vibration along the $z$ axis at
this level of approximation.

\section{Vibrations along the $y$ axis}

We first consider vibrations along the $y$ axis. The measure of the
quality of the interferometer is it's contrast, so we calculate and plot the
contrast versus the frequency of oscillations. The four-blade version returns higher contrast by compensating  with the third blade for momentum change introduced by vibration of the second blade. We expect this compensation to be most precise for low-frequency vibrations.

\subsection{Three-blade interferometer}
Assume that the neutron hits the first blade of interferometer at
the time $t=0$. We rewrite Eq.~\ref{eq:oscil} for vibrations
along the $y$ axis where $\phi$ is a random phase between the
arriving neutron and the vibrating blade.
\begin{equation}
y(t)=y_0\sin{(\omega t + \varphi)} \label{eq:oscil_y}.
\end{equation}
The velocity of the interferometer at the time $t$ is
\begin{equation}
u_y(t)=\frac{dy(t)}{dt}=y_0 \omega \cos{(\omega t + \varphi)}
\label{eq:oscil_y_u}.
\end{equation}
At time $t=0$, the velocity of the interferometer along the
$y$-coordinate is $u_y(0)=y_0 \omega \cos{\varphi}$, where $\varphi$
is random. Conservation of momentum and energy at the moment $t=0$
implies that the velocity of the transmitted neutron does not change,
while the reflected neutron's velocities are
$\vec{v}_{refl}(0+)=v_x\hat{\mathbf{x}}-(v_y-2u_y(0))\hat{\mathbf{y}}$.

The phase difference for the neutron between $path~I$ and $path~II$ is
\begin{equation}
\Delta \Phi=\Phi(path~II)-\Phi(path~I)=\frac{1}{\hbar}
\int_{path~II}{\mathbf{p}d\mathbf{s}} - \frac{1}{\hbar}
\int_{path~I}{\mathbf{p}d\mathbf{s}} \label{eq:d_phi_def},
\end{equation}
where $\mathbf{p}$ is the momentum of the neutron and $\mathbf{s}$
is the path-length vector along which the neutron is moving.

For the neutron to travel between the first two blades takes a time
$t=2L/v_x=2\tau$. The contrast depends on the total phase difference
between the paths. Notice that under these assumptions the
two paths cross the third blade at the same spot and the travel time
along these paths remains $4\tau$. So, the loss in contrast seen
in the presence of vibration is not due to the finite coherence
length of the interferometer but rather is due to the extra phase
shifts introduced by the vibrations.

Using these we find
\begin{eqnarray}
\Delta \Phi(\varphi)=16\frac{m_n}{\hbar} \tau
\left(v_y-u_y(0)\right)\left(u_y(2\tau)-u_y(0)\right)
\label{eq:d_phi_3bl2}.
\end{eqnarray}
If we assume that $u_y(t)$ is slowly varying on the scale of $2\tau$
(or $\omega \tau \ll 1$) we can approximate the expression
$\left(u_y(2\tau)-u_y(0)\right)$ as a derivative of $u_y(t)$
\begin{eqnarray}
\Delta \Phi(\varphi)=16\frac{m_n}{\hbar} \tau^2
\left(v_y-u_y(0)\right)2\left. \frac{du_y(t)}{dt}\right|_{t=\tau}
\label{eq:d_phi_3bl_deriv}.
\end{eqnarray}

The intensity at the O-detector is
\begin{equation}
I_O(\phi)=1+ \cos{\left( \Delta \Phi (\varphi) +\phi \right)}
\label{eq:int_3b_y}
\end{equation}
and depends on the random phase $\varphi$. We average
the intensity over this phase:
\begin{equation}
\overline{I_O(\phi)}=\frac{1}{2\pi}\int^{2\pi}_{0}{\left(1+ \cos{\left( \Delta
\Phi (\varphi) +\phi \right)}\right)d\varphi}
\label{eq:avg_int_3b_y}
\end{equation}

The contrast $C$ as usual is defined as
\begin{equation}
C=\frac{\max\{\overline{I(\phi)}\} - \min\{\overline{I(\phi)}\}
}{\max\{\overline{I(\phi)}\} + \min\{\overline{I(\phi)}\}}
\label{eq:contr_def},
\end{equation}
where we vary the phase $\phi$ to find the max and min.

\begin{figure}[!ht]
\includegraphics[width=1\textwidth]{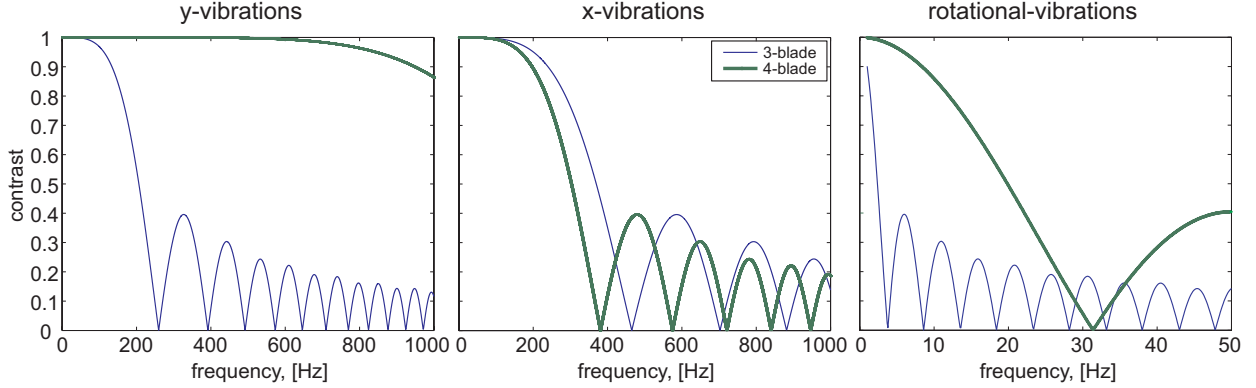}
\caption{\label{fig:vib} \textbf{A} Contrast due to vibrations along
the $y$ axis. \textbf{B} Contrast due to vibrations along the
$x$ axis. \textbf{C} Contrast due to rotational vibrations $\theta$
around $z$ axis and the center of mass. Note that the scale of the frequency
axis in \textbf{C} is different from those shown on \textbf{A} and \textbf{B} }
\end{figure}

Fig.~\ref{fig:vib}~\textbf{A} shows the dependence of the contrast
$C_y$ on the frequency of vibrations along the $y$ axis for the
three-blade interferometer. The
contrast was calculated for $L=5$~cm, a neutron velocity of
$v=2000$~m/s, and vibration amplitudes of $y_0=0.1~\mu$m. Here we
observe that the contrast starts to decrease near 100~Hz.

\subsection{Four-blade interferometer}

As in the three-blade interferometer case, we derive
expressions for the phases the neutron acquires while traveling along
the two interferometer paths. Here, the time for the neutron to
travel between the first two blades is $\tau=L/v_x$. Note that the detector corresponding to the neutron paths with equal number of reflections for the three-blade interferometer is O-detector and for the four-blade interferometer it is H-detector.

Between the first blade and the second, the phases are identical to
the three-blade interferometer case except that $2L$ changes to $L$. The phase
difference between $path~II$ and $path~I$ is
\begin{eqnarray}
\Delta \Phi(\varphi)=8\frac{m_n}{\hbar} \tau
\left(u_y(0)-v_y\right)\left(2u_y(0)-3u_y(\tau)+u_y(3\tau)\right)
\label{eq:d_phi_4bl2}.
\end{eqnarray}
Again, as for the case of three-blade interferometer, we assume that the
function $u_y(t)$ is slowly varying on the scale of $\tau$ (or
$\omega \tau \ll 1$) and we rewrite the phase change in terms of a
derivative
\begin{eqnarray}
\Delta \Phi(\varphi)&=&16\frac{m_n}{\hbar} \tau^2
\left(v_y-u_y(0)\right)\left[\left.\frac{du_y(t)}{dt}\right|_{t=\frac{\tau}{2}}-
\left.\frac{du_y(t)}{dt}\right|_{t=2\tau}\right]=\nonumber \\
&&16\frac{m_n}{\hbar}\tau^3
\left(v_y-u_y(0)\right)\frac{3}{2}\left.\frac{du_y^2(t)}{d^2t}\right|_{t=\frac{5\tau}{4}}
\label{eq:d_phi_4bl_der}.
\end{eqnarray}

Notice that the linear term drops out. This is the source of the protection against
vibrations. The contrast comparison we
make is the O-beam of the three-blades interferometer to the H-beam of
the four-blade interferometer. The intensity at the
H-detector is
\begin{equation}
I_H(\phi)=1+ \cos{\left( \Delta \Phi (\varphi) +\phi \right)}
\label{eq:int_4b_y}
\end{equation}
and depends on the random phase $\varphi$ of vibration. Again we
average the intensity over this random phase:
\begin{equation}
\overline{I_H(\phi)}=\frac{1}{2\pi} \int^{2\pi}_{0}{\left(1+ \cos{\left( \Delta
\Phi (\varphi) +\phi \right)}\right)d\varphi}
\label{eq:avg_int_4b_y}.
\end{equation}

We obtain the contrast using Eq.~\ref{eq:contr_def}. In
Fig.~\ref{fig:vib}~\textbf{A} we plot the frequency dependence of
the contrast for the four-blade interferometer. Notice that
$paths~I$ and $II$ cross the fourth blade at the same spot. From
Fig.~\ref{fig:vib}~\textbf{A} we clearly see that the four-blade
interferometer is predicted to be much less sensitive to
$y$ vibrations.

\section{Vibrations along the $x$ axis}

In the case of vibrations along the $x$ axis the momentum of the neutron
is not modified (see Fig.~\ref{fig:inter_sch}). However the path
length changes depending on the phase $\varphi$ of the oscillations
at $t=0$.

Here the contrast is primarily
limited by the neutron coherence length of $1/ \Delta k$.
The four-blade geometry does not protect against this and indeed the influence of a finite coherence length is slightly worse due to the noise being introduced since the intervals between blades are changed by the vibrations.  In the three-blade case there are two such intervals, while in the four-blade case there are three.  So with finite $\Delta k$ the more blades there are, the worse the noise is.  However, since the acceptance of the NI is small, $\Delta k$ is small and this contribution to the noise is small.

\begin{figure}[!ht]
\centering
\includegraphics[width=0.7\textwidth]{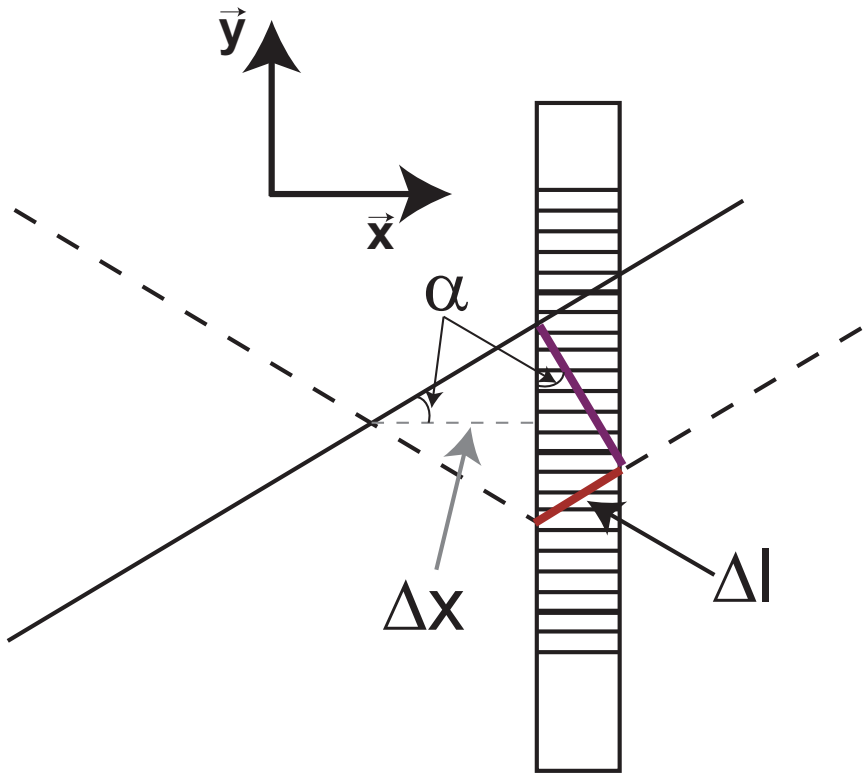}
\caption{\label{fig:dx_df_sch} A schematic diagram of the neutron
arriving to the third blade. Due to the crystal movement, the paths
of the neutron will not recombined at the ideal point but at $\Delta
x$ away from the third blade.}
\end{figure}

Here we calculate just the contribution to the loss in contrast from the phase shift introduced by vibrations along $x$ axis.
The vibrations along the $x$ axis are
\begin{equation}
x(t)=x_0\sin{(\omega t + \varphi)} \label{eq:oscil_x}.
\end{equation}
In this case the phase shift is due to the paths crossing at a point
displaced from ideal as shown in Fig.~\ref{fig:dx_df_sch}.  Once
we find $\Delta x$ for each interferometer, then $\Delta l=2\Delta
x \tan{\alpha}\sin{\alpha}$ and the phase difference is
\begin{equation}
\Delta \Phi(\varphi)=\frac{m_n}{\hbar}v \Delta l
\label{eq:phase_dif_x_def}.
\end{equation}

For the three-blade interferometer, we have $\tau=L/v_x$ and
\begin{equation}
\Delta x\approx x(4\tau)-2x(2\tau)+x(0) \label{eq:delta_x_3bl},
\end{equation}
where we neglect distance $x(2\tau)-x(0)$ ($<1\mu$m) in comparison
with $L$ ($>1$cm). For the four-blade interferometer
\begin{equation}
\Delta x=x(4\tau)-4x(\tau)+3x(0) \label{eq:delta_x_4bl}.
\end{equation}
Using these $\Delta x$ we can get $\Delta \Phi$, substitute obtained
$\Delta \Phi$ to find intensities, and average intensities over
different $\varphi$ to obtain the contrasts.

Figure~\ref{fig:vib}~\textbf{B} shows the contrast dependence on the
frequency of vibrations along the $x$ axis.
This should be viewed as an upper bound on the predicted contrast. The message is that vibrations along $x$ axis are not very important and even though the four-blade design is somewhat more sensitive to them than is the three-blade design. In the final analysis the four-blade design is predicted to have better overall performance.

\section{Rotation around the $z$ axis}

These vibrations are expected to be most limiting since the neutron
interferometer has such a small acceptance angle. In the case of
rotational vibrations, we rewrite the oscillation in terms of the
angle $\theta$ around the $z$ axis
\begin{equation}
\theta(t)=\theta_0\sin{(\omega t + \varphi)} \label{eq:oscil_r}.
\end{equation}
For small angles, rotational vibrations can be considered as
translational vibrations, i.e. $\Delta r=r*\theta$, where $r$ is the
distance from the blade to the center of rotation.

\subsection{Three-blade interferometer}

In the three-blade interferometer, the center of rotation is also the
center of mass and the center of the middle blade. For the point
[see Fig.~\ref{fig:4bl_sch}~\textbf{A})] where the neutron path
crosses the blades, the rotational vibrations can be modeled as
vibration along the $y$ axis for the first and last blade and along
the $x$ axis for the path crossing the middle blade. In this case,
the interaction with the middle blade does not change the velocity
of the neutron. At the first blade we have a change in the momentum
of the reflected beam and no change for the transmitted neutrons
\begin{equation}
\vec{v}_{path~I}(t=0+)= v_x \mathbf{\hat{x}} +v_y \mathbf{\hat{y}}
\label{eq:3b1_v_t},
\end{equation}
\begin{equation}
\vec{v}_{path~II}(t=0+)= v_x \mathbf{\hat{x}} + (-v_y+2u1_y(0))
\mathbf{\hat{y}} \label{eq:3b2_v_t},
\end{equation}
where $u1_y(t)=2L\theta_0\sin{(\omega t + \varphi)}$ velocity of the
first blade in the $\mathbf{\hat{y}}$ direction.

The phase difference between the two paths is
\begin{eqnarray}
\Delta
\Phi(\varphi)=\Phi(path~II)-\Phi(path~I)=\frac{m_n}{\hbar}\left[
|\vec{v}_{path~I}|^2 - |\vec{v}_{path~II}|^2 \right]4\tau =\nonumber
\\
\frac{m_n}{\hbar}\left[(v_x^2+v_y^2-4u1_y(0)v_y + (2u1_y(0))^2-
v_x^2-v_y^2\right]4\tau = \nonumber \\
8\frac{m_n}{\hbar} L\theta_0\omega\sin{\varphi}
\left[2L\theta_0\omega\sin{\varphi}-v_y\right]4\tau
\label{eq:d_phi_3bl_r},
\end{eqnarray}
where $\tau=L/v_x$.

Substituting this difference in phase into
Eq.~\ref{eq:int_3b_y} for the O-beam intensity and averaging,
we find the frequency dependence of the contrast~\ref{eq:contr_def}. Fig.~\ref{fig:vib}~\textbf{C} shows this
contrast. As
an amplitude of vibrations $\theta_0$ we used $1~\mu$rad. The value for the amplitude comes from the current NIST setup, where a vibration isolation table is controlled to this level~\cite{arif_inter}.

\subsection{Four-blade interferometer}

In the four-blade interferometer the center of rotation and the center
of mass coincide between the blades. For points [see
Fig.~\ref{fig:4bl_sch}~\textbf{B})] where the neutron path crosses
the blades, the rotational vibrations are modeled as vibrations
along the $y$ axis. As in the three-blade case the $v_x$ component of
the neutron velocity does not change. The velocities are modified
as follows.

For $path~I$:
\begin{eqnarray}
v_y(I:1\rightarrow 2)&=& v_y, \\
v_y(I:2\rightarrow 3)&=& -v_y+2\sqrt{L^2+(v_y\tau)^2}\theta_0 \omega \cos{(\omega\tau+\varphi)}, \\
v_y(I:3\rightarrow 4)&=& v_y-2\sqrt{L^2+(v_y\tau)^2}\theta_0
\omega\left[ \cos{(\omega\tau+\varphi)} + \cos{(\omega
3\tau+\varphi)} \right] \label{eq:v_4b1_r},
\end{eqnarray}
where $\tau=L/v_x$ and the sign of the last cosine term is positive
because the oscillations of the second and the third blades
have a $\pi$ phase-shift difference.

For $path~II$:
\begin{eqnarray}
v_y(II:1\rightarrow 2)&=& v_y+2\theta_0 \omega 2L\cos{\phi}, \\
v_y(II:2\rightarrow 3)&=& -v_y(II:1\rightarrow 2)+2\sqrt{L^2+(v_y\tau)^2}\theta_0 \omega \cos{(\omega\tau+\varphi)}, \\
v_y(II:3\rightarrow 4)&=& -v_y(II:2\rightarrow 3) -
2\sqrt{L^2+(v_y\tau)^2} \theta_0 \omega \cos{(\omega 3\tau+\varphi)}
\label{eq:v_4b2_r}.
\end{eqnarray}

The phases along each path are as follows.

For $path~I$:
\begin{eqnarray}
\Phi(I:1\rightarrow 2)&=& \frac{m_n}{\hbar}v^2\tau, \\
\Phi(I:2\rightarrow 3)&=& \frac{m_n}{\hbar}(v_x^2+v_y(I:2\rightarrow 3)^2)2\tau, \\
\Phi(I:2\rightarrow 3)&=& \frac{m_n}{\hbar}(v_x^2+v_y(I:3\rightarrow
4)^2)\tau \label{eq:phase_4b1_r}.
\end{eqnarray}
For $path~II$:
\begin{eqnarray}
\Phi(II:1\rightarrow 2)&=& \frac{m_n}{\hbar}(v_x^2+v_y(II:1\rightarrow 2)^2)\tau, \\
\Phi(II:2\rightarrow 3)&=& \frac{m_n}{\hbar}(v_x^2+v_y(II:2\rightarrow 3)^2)2\tau, \\
\Phi(II:2\rightarrow 3)&=&
\frac{m_n}{\hbar}(v_x^2+v_y(II:3\rightarrow 4)^2)\tau
\label{eq:phase_4b2_r}.
\end{eqnarray}

The phase difference is
\begin{eqnarray}
\Delta \Phi(\varphi)=\Phi(II:1\rightarrow2)+
\Phi(II:2\rightarrow3)+ \Phi(II:3\rightarrow4) - \nonumber \\
\left(\Phi(I:1\rightarrow2)+ \Phi(I:2\rightarrow3) +
\Phi(I:3\rightarrow4)\right) \label{eq:d_phi_4bl_r}.
\end{eqnarray}

As before, we can find the $I_H$ intensity at the H-detector,
average it over the random phase $\varphi$, and extract the
contrast. This contrast dependence on the frequency of rotational
vibrations is plotted in
Fig.~\ref{fig:vib}~\textbf{C}. We see that for these rotations the
four-blade interferometer design is significantly more robust than
the three-blade. Notice that for rotational vibrations the first moment of the loss of contrast is not equal to zero; all vibrations contribute to the loss of contrast.

\section{Conclusion}

Our model reconstructs the situation that is normally seen in neutron
interferometry.
Vibrations we use in our simulations (with amplitude $10^{-7}$m in
translation) produce changes in the incident angle of the neutron of
much less than the acceptance angle of the crystal ($\ll 5\times
10^{-6}$rad) and of a similar order for the 50Hz frequency range previously
measured~\cite{arif_inter}. In order to exceed the
acceptance angle the amplitude of vibrations would have to be bigger than
50$\mu$m.

Note that small angle vibrations around  the $x$ axis will
be similar to the translational vibrations along the $y$ axis, and
small angle vibrations around $y$ will be similar to the
translational vibrations along $x$. As mentioned before, vibrations along the $z$ axis do not influence the contrast.

Taken together these results  bring us to the four-blade
experimental geometry for neutron interferometer. Although the new four-blade design leads to a loss of half of the neutron intensity we can make up this loss with a more robust, small system that can be moved closer to the beam break. We thus regain and even increase the final neutron intensity at the detectors. In order to achieve high contrast the proposed four-blade interferometer system still requires excellent temperature stability~\cite{pushin_prl_coherence:250404}.

The robust nature of the four-blade interferometer can also be understood  as a result of dynamical decoupling  or decoherence-free subspace (DFS).  The dynamical decouping analogy~\cite{Lloyd_bang_bang_PhysRevA.58.2733,Lloyd_bang_bang_PhysRevLett.82.2417} is easiest to see; the third set of blades act to undo the change in momentum introduced by the second blades.  Provided that the noise (motion of the NI) is the same when the neutron encounters both the second and third blades, then the momentum error is completely removed.

\begin{figure}[t]
\centering
\includegraphics[width=0.8\textwidth]{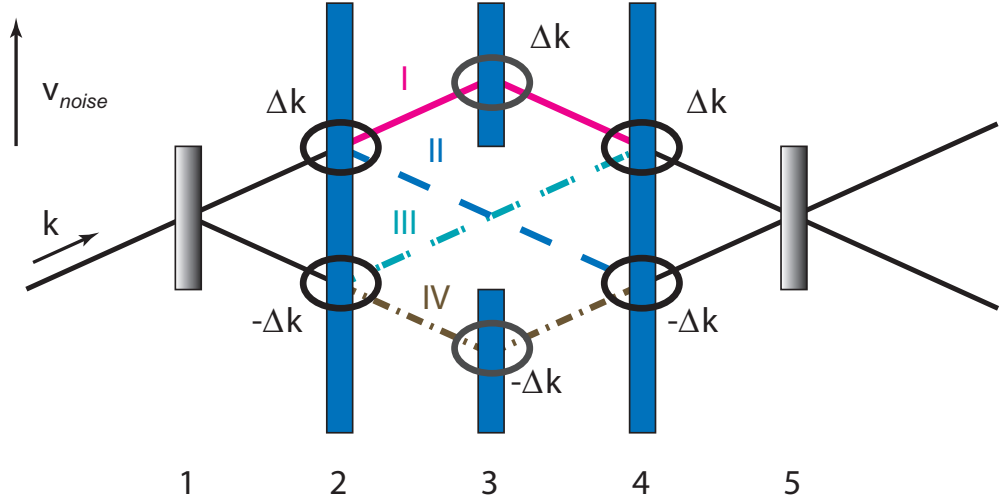}
\caption{\label{fig:dfs} A schematic representation of the robust four-blade interferometer imbedded in a five-blade interferometer. The overall interferometer operates over SU(4) and can be thought of as a tensor product space of two qubits. A simple abstraction of the noise generators is $\pm \Delta k$ at each internal reflection (there is no noise at the first and final blades).\\
When we look over the entire path:
two paths have noise ($I$ and $IV$), and
two paths do not have noise ($II$ and $III$).
The noise generator is  proportional to $sgn(\vec{k} \cdot \vec{v}_{noise})  \Delta k$ for each reflection internal to NI.
}
\end{figure}

The DFS~\cite{DFS_zanardi,DFS_duan} picture can be seen if two additional paths are considered.  now we have four paths and an interferometer that acts over SU(4); see Fig.~~\ref{fig:dfs}. The outer two paths are sensitive to the noise and the inner two are isolated from it.  We are presently installing a five-blade interferometer to test these predictions.

\section{Acknowledgement}
Support provided by NIST is gratefully acknowledged. The authors are grateful for
discussions with D. L. Jacobson, R. Laflamme, B. Levi, C. Ramanathan. Discussion with
S.A. Werner about an earlier idea of a four-blade interferometer for use in gravity and spin-rotation \cite{buras.moving.lattices} experiments is gratefully appreciated.


\begin{thebibliography}{14}
\expandafter\ifx\csname natexlab\endcsname\relax\def\natexlab#1{#1}\fi
\expandafter\ifx\csname bibnamefont\endcsname\relax
  \def\bibnamefont#1{#1}\fi
\expandafter\ifx\csname bibfnamefont\endcsname\relax
  \def\bibfnamefont#1{#1}\fi
\expandafter\ifx\csname citenamefont\endcsname\relax
  \def\citenamefont#1{#1}\fi
\expandafter\ifx\csname url\endcsname\relax
  \def\url#1{\texttt{#1}}\fi
\expandafter\ifx\csname urlprefix\endcsname\relax\def\urlprefix{URL }\fi
\providecommand{\bibinfo}[2]{#2}
\providecommand{\eprint}[2][]{\url{#2}}

\bibitem[{\citenamefont{Rauch and Werner}(2000)}]{ni_book}
\bibinfo{author}{\bibfnamefont{H.}~\bibnamefont{Rauch}} \bibnamefont{and}
  \bibinfo{author}{\bibfnamefont{S.~A.} \bibnamefont{Werner}},
  \emph{\bibinfo{title}{Neutron Interferometry}} (\bibinfo{publisher}{Oxford
  University Press}, \bibinfo{year}{2000}).

\bibitem[{\citenamefont{Rauch et~al.}(1974)\citenamefont{Rauch, Treimer, and
  Bonse}}]{rauch1int1974}
\bibinfo{author}{\bibfnamefont{H.}~\bibnamefont{Rauch}},
  \bibinfo{author}{\bibfnamefont{W.}~\bibnamefont{Treimer}}, \bibnamefont{and}
  \bibinfo{author}{\bibfnamefont{U.}~\bibnamefont{Bonse}},
  \bibinfo{journal}{Phys. Lett. A} \textbf{\bibinfo{volume}{47}},
  \bibinfo{pages}{369} (\bibinfo{year}{1974}).

\bibitem[{\citenamefont{Bauspiess et~al.}(1974)\citenamefont{Bauspiess, Bonse,
  Rauch, and Treimer}}]{bauspiess1int1974}
\bibinfo{author}{\bibfnamefont{W.}~\bibnamefont{Bauspiess}},
  \bibinfo{author}{\bibfnamefont{U.}~\bibnamefont{Bonse}},
  \bibinfo{author}{\bibfnamefont{H.}~\bibnamefont{Rauch}}, \bibnamefont{and}
  \bibinfo{author}{\bibfnamefont{W.}~\bibnamefont{Treimer}},
  \bibinfo{journal}{Z. Physik} \textbf{\bibinfo{volume}{271}},
  \bibinfo{pages}{177} (\bibinfo{year}{1974}).

\bibitem[{\citenamefont{Shimony and Feshbach}(1987)}]{shull.vibrations.book}
\bibinfo{author}{\bibfnamefont{A.}~\bibnamefont{Shimony}} \bibnamefont{and}
  \bibinfo{author}{\bibfnamefont{H.}~\bibnamefont{Feshbach}},
  \emph{\bibinfo{title}{Physics As Natural Philosophy}}
  (\bibinfo{publisher}{MIT Press}, \bibinfo{year}{1987}).

\bibitem[{\citenamefont{Arif et~al.}(1994)\citenamefont{Arif, Brown, Greene,
  Clothier, and Littrell}}]{arif_inter}
\bibinfo{author}{\bibfnamefont{M.}~\bibnamefont{Arif}},
  \bibinfo{author}{\bibfnamefont{D.~E.} \bibnamefont{Brown}},
  \bibinfo{author}{\bibfnamefont{G.~L.} \bibnamefont{Greene}},
  \bibinfo{author}{\bibfnamefont{R.}~\bibnamefont{Clothier}}, \bibnamefont{and}
  \bibinfo{author}{\bibfnamefont{K.}~\bibnamefont{Littrell}},
  \bibinfo{journal}{Proc. SPIE Int. Soc. Opt. Eng.}
  \textbf{\bibinfo{volume}{2264}}, \bibinfo{pages}{20} (\bibinfo{year}{1994}).

\bibitem[{\citenamefont{Shull and Gingrich}(1964)}]{shull.moving.xtal:678}
\bibinfo{author}{\bibfnamefont{C.~G.} \bibnamefont{Shull}} \bibnamefont{and}
  \bibinfo{author}{\bibfnamefont{N.~S.} \bibnamefont{Gingrich}},
  \bibinfo{journal}{Journal of Applied Physics} \textbf{\bibinfo{volume}{35}},
  \bibinfo{pages}{678} (\bibinfo{year}{1964}),
  \urlprefix\url{http://link.aip.org/link/?JAP/35/678/1}.

\bibitem[{\citenamefont{Buras and Giebultowicz}(1972)}]{buras.moving.lattices}
\bibinfo{author}{\bibfnamefont{B.}~\bibnamefont{Buras}} \bibnamefont{and}
  \bibinfo{author}{\bibfnamefont{T.}~\bibnamefont{Giebultowicz}},
  \bibinfo{journal}{Acta Cryst. A} \textbf{\bibinfo{volume}{28}},
  \bibinfo{pages}{151} (\bibinfo{year}{1972}).

\bibitem[{\citenamefont{Dresden and
  Yang}(1979)}]{dresden.sagnac.effect.PhysRevD.20.1846}
\bibinfo{author}{\bibfnamefont{M.}~\bibnamefont{Dresden}} \bibnamefont{and}
  \bibinfo{author}{\bibfnamefont{C.~N.} \bibnamefont{Yang}},
  \bibinfo{journal}{Phys. Rev. D} \textbf{\bibinfo{volume}{20}},
  \bibinfo{pages}{1846} (\bibinfo{year}{1979}).

\bibitem[{\citenamefont{Mashhoon}(1988)}]{mashhoon.PhysRevLett.61.2639}
\bibinfo{author}{\bibfnamefont{B.}~\bibnamefont{Mashhoon}},
  \bibinfo{journal}{Phys. Rev. Lett.} \textbf{\bibinfo{volume}{61}},
  \bibinfo{pages}{2639} (\bibinfo{year}{1988}).

\bibitem[{\citenamefont{Pushin et~al.}(2008)\citenamefont{Pushin, Arif, Huber,
  and Cory}}]{pushin_prl_coherence:250404}
\bibinfo{author}{\bibfnamefont{D.~A.} \bibnamefont{Pushin}},
  \bibinfo{author}{\bibfnamefont{M.}~\bibnamefont{Arif}},
  \bibinfo{author}{\bibfnamefont{M.~G.} \bibnamefont{Huber}}, \bibnamefont{and}
  \bibinfo{author}{\bibfnamefont{D.~G.} \bibnamefont{Cory}},
  \bibinfo{journal}{Physical Review Letters} \textbf{\bibinfo{volume}{100}},
  \bibinfo{eid}{250404} (pages~\bibinfo{numpages}{4}) (\bibinfo{year}{2008}),
  \urlprefix\url{http://link.aps.org/abstract/PRL/v100/e250404}.

\bibitem[{\citenamefont{Viola and
  Lloyd}(1998)}]{Lloyd_bang_bang_PhysRevA.58.2733}
\bibinfo{author}{\bibfnamefont{L.}~\bibnamefont{Viola}} \bibnamefont{and}
  \bibinfo{author}{\bibfnamefont{S.}~\bibnamefont{Lloyd}},
  \bibinfo{journal}{Phys. Rev. A} \textbf{\bibinfo{volume}{58}},
  \bibinfo{pages}{2733} (\bibinfo{year}{1998}).

\bibitem[{\citenamefont{Viola et~al.}(1999)\citenamefont{Viola, Knill, and
  Lloyd}}]{Lloyd_bang_bang_PhysRevLett.82.2417}
\bibinfo{author}{\bibfnamefont{L.}~\bibnamefont{Viola}},
  \bibinfo{author}{\bibfnamefont{E.}~\bibnamefont{Knill}}, \bibnamefont{and}
  \bibinfo{author}{\bibfnamefont{S.}~\bibnamefont{Lloyd}},
  \bibinfo{journal}{Phys. Rev. Lett.} \textbf{\bibinfo{volume}{82}},
  \bibinfo{pages}{2417} (\bibinfo{year}{1999}).

\bibitem[{\citenamefont{Zanardi and Rasetti}(1997)}]{DFS_zanardi}
\bibinfo{author}{\bibfnamefont{P.}~\bibnamefont{Zanardi}} \bibnamefont{and}
  \bibinfo{author}{\bibfnamefont{M.}~\bibnamefont{Rasetti}},
  \bibinfo{journal}{Phys. Rev. Lett.} \textbf{\bibinfo{volume}{79}},
  \bibinfo{pages}{3306} (\bibinfo{year}{1997}).

\bibitem[{\citenamefont{Duan and Guo}(1997)}]{DFS_duan}
\bibinfo{author}{\bibfnamefont{L.-M.} \bibnamefont{Duan}} \bibnamefont{and}
  \bibinfo{author}{\bibfnamefont{G.-C.} \bibnamefont{Guo}},
  \bibinfo{journal}{Phys. Rev. Lett.} \textbf{\bibinfo{volume}{79}},
  \bibinfo{pages}{1953} (\bibinfo{year}{1997}).

\end{thebibliography}

\end{document}